\documentclass[conference]{IEEEtran}

\usepackage[letterpaper,
  left=0.75in,
  right=0.75in,
  top=0.8in,
  bottom=1in]{geometry}

\usepackage[T1]{fontenc}
\usepackage{lmodern}

\normalsize

\normalsize

\ifCLASSINFOpdf

\else

\fi

\usepackage{algorithm}
\usepackage{algorithmic}

\usepackage{amsmath}
\usepackage{multirow}
\usepackage{graphicx}
\usepackage{epstopdf}
\usepackage{color}
\usepackage{soul}
\usepackage{placeins}
\usepackage{float}
\usepackage{tabularx,colortbl}
\usepackage{url}

\makeatletter
\def\@IEEEbibitemmax{2} 
\makeatother

\begin{document}

\newcommand{\MYfooter}{\smash{
\hfil\parbox[t][\height][t]{\textwidth}{\centering
\thepage}\hfil\hbox{}}}

\makeatletter

\makeatother
\pagestyle{headings}
\addtolength{\footskip}{0\baselineskip}
\addtolength{\textheight}{-1\baselineskip}

%
\title{Coexistence of 5G NR and Wi‑Fi 6E/7 at 6 GHz: Experimental Interference Measurements}

\author{
\IEEEauthorblockN{
Rafik Zitouni\textsuperscript{$\star$},
Demos Serghiou\textsuperscript{$\star$},
Ali Dagdeviren\textsuperscript{$\diamond$},
Tajinder Randhawa\textsuperscript{$\diamond$},\\
Edwards Udean\textsuperscript{$\star$},
Hanli Dong\textsuperscript{$\star$},
Riccardo Pozza\textsuperscript{$\star$},
Rahim Tafazolli\textsuperscript{$\star$},}
\IEEEauthorblockA{
\textsuperscript{$\star$}6GIC, Institute for Communication Systems (ICS), University of Surrey, Guildford, U.K. \\
\textsuperscript{$\diamond$}ArteLabs, Morden, London, U.K. \\
Email: \textsuperscript{$\star$}\{r.zitouni, demos.serghiou\}@surrey.ac.uk,
\textsuperscript{$\diamond$}\{ali.dagdeviren, trandhawa\}@artelabs.co.uk}}

\maketitle


\begin{abstract}

This paper presents the first conducted-interference measurements of a
commercial Very Low Power (VLP) Wi-Fi 6E/7 device into both the gNB uplink
and UE downlink receiver chains of a live 5G New Radio (NR) system, using a
complete O-RAN/SDR stack with 5G core in band~n102 (6\,GHz). We use a
Software-Defined Radio (SDR) testbed built on OpenAirInterface with
band~n102 support (40\,MHz, 30\,kHz Subcarrier Spacing). We sweep the
injected Wi-Fi power and record throughput, block error rate, and
signal-to-noise ratio on both the gNB uplink and UE downlink paths. Neither
receiver shows measurable degradation below $-75$\,dBm. Above this
threshold, performance degrades progressively. The UE is more resilient at
lower data rates and unaffected by beacon-only transmissions. A complementary
link-budget analysis maps these measured thresholds to equivalent
VLP-to-victim distances. These distances fall well inside the 545--685\,m
Listen Before Talk (LBT) exclusion zone, confirming that a compliant VLP
device would vacate the channel before its emissions could harm either
receiver.

\end{abstract}

\vspace{0.2cm}

\IEEEpeerreviewmaketitle

\section{Introduction}

The opening of the 6\,GHz band for unlicensed access is one of the most significant spectrum-policy developments of the past decade. The FCC allocated 1.2\,GHz of new spectrum~\cite{b1}, while European regulators opened the lower 500\,MHz~\cite{b2}. Two radio access technologies (RATs) are expected to share these bands: Wi-Fi~6E/7 (IEEE~802.11ax/be)~\cite{b3}, operating as Very Low Power (VLP) devices, and 5G New Radio Unlicensed (NR-U)~\cite{b4} as part of Mobile/Fixed Communications Networks (MFCN). The terms ``VLP'' and ``Wi-Fi'' are used interchangeably throughout this paper.

Unlike the 5\,GHz bands, the 6\,GHz range has no pre-existing unlicensed deployments. Coexistence mechanisms can therefore be designed from scratch. Both RATs employ Listen-Before-Talk (LBT) protocols derived from Carrier-Sense Multiple Access with Collision Avoidance (CSMA/CA), yet differ in scheduling architecture, transmission-opportunity duration, and uplink access mechanisms. Rigorous experimental characterisation of their interaction is therefore essential.

The paper is organised as follows. Section~\ref{sec_related_cont} surveys related work and states our contributions. Section~\ref{sec_analytical_model} develops the analytical coexistence model. Sections~\ref{sec_experimental_setup} and~\ref{sec_measurements} detail the experimental setup and measurement results. Section~\ref{sec_discussion} discusses the implications, and Section VII concludes the paper.

\section{Related work and paper contributions}
\label{sec_related_cont}
Naik et al.~\cite{b5} used stochastic geometry to derive closed-form transmission success probabilities for Wi-Fi~6E and NR-U sharing the 6\,GHz band. The same authors identified the inter-RAT Energy-Detection (ED) threshold as the most consequential unresolved parameter, noting that no consensus exists on whether energy detection, preamble detection, or a hybrid scheme best enables fair coexistence~\cite{b6}. Keshtiarast and Petrova~\cite{b10} extended this with an ns-3 simulation of Wi-Fi~6E access points and 5G NR-U gNBs in dense residential deployments, finding that ED-threshold selection matters more for fairness than Maximum Channel Occupancy Time (MCOT) alone. However, both studies remain simulation-based and downlink-oriented. Neither characterises real Wi-Fi~6E/7 devices or gNB UL receiver chains under controlled interference.

Adjacent-channel and legacy-service protection scenarios have also been studied. Pastukh et al.~\cite{b7} used Monte Carlo simulation to quantify Signal-to-Interference-plus-Noise Ratio (SINR) degradation in 5G NR DL throughput from indoor Wi-Fi in the neighbouring 5925--6425\,MHz band, reporting losses of up to 22.5\% under line-of-sight conditions. Yoza-Mitsuishi et al.~\cite{b8} analysed aggregate interference from Wi-Fi and 5G NR-U devices into fixed-satellite service uplinks across the United States. The authors in~\cite{b11} conducted an extensive measurement campaign on a dense indoor Wi-Fi~6E network; their focus, however, is leakage from indoor Wi-Fi~6E into incumbent fixed links rather than co-channel interference into a gNB UL receiver.

Collectively, these works~\cite{b5,b6,b7,b8,b10,b11} provide analytical, simulation, and measurement foundations for 6\,GHz coexistence, yet a common gap remains: no study offers hardware-in-the-loop validation of Wi-Fi~6E/7 interference into both gNB UL and UE DL receiver chains. The behaviour of real LBT MAC implementations under controlled co-channel interference is uncharacterised, and no end-to-end 5G~NR setup with a full core network has been tested against a live VLP device.

The contributions of this paper are:
\begin{itemize}
  \item First conducted-interference measurements of a commercial VLP Wi-Fi~6E/7 device into a 5G~NR system (band~n102) using a complete O-RAN/SDR stack with 5G core, covering both the gNB UL and UE DL receiver chains.
  \item Per-dB characterisation of the degradation starting point at each receiver via SNR, BLER, and throughput sweeps.
  \item A link-budget model that derives two distances: the harm radius (how far the VLP signal can travel before causing degradation) and the LBT exclusion zone (how far the gNB signal forces the VLP to defer). The exclusion zone is several times larger than the harm radius, so the VLP vacates the channel before it can degrade either receiver.
\end{itemize}

\section{Analytical Coexistence Model} \label{sec_analytical_model}
This section derives the spatial relationship between two quantities: (i)~the \emph{LBT exclusion radius}~$d_{\text{LBT}}$, within which a VLP device senses gNB broadcast power above its Energy Detection (ED) threshold and defers, and (ii)~the \emph{harm radius}~$d_{\text{harm}}$, the maximum distance at which VLP emissions reach the $-75$\,dBm degradation threshold measured in Section~\ref{sec_measurements}. Coexistence holds when $d_{\text{harm}} \ll d_{\text{LBT}}$.

The dominant interference path runs from a Wi-Fi~6E/7 VLP access point toward the 5G~NR gNB uplink and UE downlink receivers. ECC Report~366~\cite{b9} studies shared use of the 5925--7125\,MHz band under this geometry. The relevant 3GPP bands (TS~38.104~\cite{b13}) are n102 (5925--6425\,MHz) and n104 (6425--7125\,MHz); we target n102, though the analysis extends to n104 with $\approx 1$\,dB additional propagation loss at 6775\,MHz.

We model received power with a log-distance path-loss model consistent with~\cite{b5,b6,b7,b8,b10}, augmented by an explicit excess-loss term $L_{\text{EX}}$ that captures non-free-space propagation:
\begin{equation}
  P^{\text{RX}}(d) = P^{\text{TX}} - L_0 - 10\gamma \log_{10}\!\left(\frac{d}{d_0}\right) - L_{\text{EX}},
  \label{equation:model}
\end{equation}
where $P^{\text{TX}}$ is the transmit Equivalent Isotropically Radiated Power (EIRP), $L_0 = 20\log_{10}(4\pi f/c) = 47.98$\,dB is the free-space path loss (FSPL) at $d_0 = 1$\,m for $f = 5995$\,MHz and $c = 3 \times 10^8$\,m/s, and $\gamma = 2$ is the free-space reference exponent. The urban propagation environment between the elevated macro gNB and street-level VLP is captured through $L_{\text{EX}}$ rather than an inflated exponent, following the agreed methodology of ECC Report~366~\cite{b9}, which specifies Recommendation ITU-R~P.1411-12~\cite{b14} for the WAS/RLAN$\leftrightarrow$MFCN path and Recommendation ITU-R~P.2109-2~\cite{b15} for outdoor-to-indoor coupling of indoor VLP devices.

\textbf{Deployment power.} A VLP device triggers its LBT on the always-on broadcast/SSB signalling, which is radiated across the sector rather than beamformed to a served UE; the governing quantity is therefore the sector broadcast EIRP, not a location-specific beamformed value. Per ECC Report~366~\cite{b9} (\S2.1.3.1), the representative macro-cell EIRP is 78--82\,dBm/100\,MHz; we adopt the sharing-study baseline of 73\,dBm/100\,MHz (Study~C10 of~\cite{b9} uses 70\,dBm/100\,MHz; up to 83\,dBm/100\,MHz is used there for SSB-detection assessments). Scaled to the VLP's 20\,MHz sensing bandwidth, this is $73 - 10\log_{10}(100/20) = 66$\,dBm/20\,MHz.

\textbf{Propagation.} For the VLP$\leftrightarrow$gNB detection path over the $0.5$--$0.7$\,km exclusion range at 6\,GHz, the ITU-R~P.1411-12 urban-NLoS model~\cite{b14} contributes a median excess loss of $L_{\text{EX}} \approx 23$\,dB above free space. For the short-range harm path (VLP$\leftrightarrow$victim) we set $L_{\text{EX}} = 0$, retaining free-space propagation as a conservative upper bound on the harm distance---permitted for the co-located link by ECC Report~366, Note~2, and understating neither the harm reach nor overstating the exclusion zone. Both choices bias the analysis toward the \emph{smallest} spatial margin.

Table~\ref{tab:coexistence_params} lists the system parameters.

\begin{table}[!t]
\centering
\caption{Wi-Fi 6E/7 (VLP) and 5G NR coexistence parameters}
\label{tab:coexistence_params}
\footnotesize
\renewcommand{\arraystretch}{1.2}
\setlength{\tabcolsep}{3pt}
\begin{tabular}{|p{1.9cm}|p{1.9cm}|p{2.0cm}|p{1.6cm}|}
\hline
\textbf{Parameter} & \textbf{Wi-Fi 6E/7 (VLP)} & \textbf{5G gNB} & \textbf{5G UE} \\
\hline
Operating band & 6\,GHz VLP & n102  & n102 \\
\hline
Centre frequency & 5995\,MHz & 5995.2\,MHz & 5995.2\,MHz \\
\hline
Channel bandwidth & 80 / 160\,MHz & 40\,MHz (106 RBs) & 40\,MHz (106 RBs) \\
\hline
Max transmit power & 14\,dBm EIRP & 73\,dBm/100\,MHz (66\,dBm/20\,MHz) & 20\,dBm (pMax) \\
\hline
CCA / ED threshold & $-62$\,dBm / 20\,MHz & N/A & N/A \\
\hline
Propagation & \multicolumn{2}{c|}{P.1411-12 urban NLoS (VLP$\leftrightarrow$gNB)} & FSPL \\
\hline
Reference clock & Internal & 10\,MHz (White Rabbit) & 10\,MHz (WR) \\
\hline
Duty cycle & 3--5\% & DL-heavy (104:38 sym) & Same as gNB \\
\hline
\end{tabular}
\end{table}

\subsection{LBT Exclusion Zone} \label{subsec:lbt_zone}

The LBT exclusion zone is the region around the gNB within which a VLP device senses NR broadcast power above its $-62$\,dBm ED threshold and defers. Setting $P^{\text{rx}}(d) = \theta_{\text{ED}}$ and inverting Equation~(\ref{equation:model}):
\begin{equation}
  d_{\text{LBT}} = d_0 \cdot 10^{\,\frac{P^{\text{TX}} - L_0 - L_{\text{EX}} - \theta_{\text{ED}}}{10\gamma}}.
  \label{eq:d_lbt}
\end{equation}

\textbf{Fully loaded} ($P^{\text{TX}} = 66$\,dBm/20\,MHz, all RBs occupied; $L_{\text{EX}} = 23.3$\,dB):
\begin{equation}
  d_{\text{LBT}} = 10^{\frac{66 - 47.98 - 23.3 - (-62)}{20}} = 10^{2.836} \approx \mathbf{685\text{\,m}}.
  \label{eq:d_lbt_full}
\end{equation}

\textbf{SSB + reference signals} (33 active RBs in 20\,MHz; effective $P^{\text{TX}} = 66 - 10\log_{10}(53/33) = 63.9$\,dBm; $L_{\text{EX}} = 23.2$\,dB):
\begin{equation}
  d_{\text{LBT,SSB}} = 10^{\frac{63.9 - 47.98 - 23.2 - (-62)}{20}} = 10^{2.736} \approx \mathbf{545\text{\,m}}.
  \label{eq:d_lbt_ssb}
\end{equation}

\subsection{VLP Harm Radius} \label{subsec:harm_radius}

The harm radius is the maximum distance at which a VLP at 14\,dBm EIRP produces the measured $-75$\,dBm degradation onset at the victim, under conservative free-space propagation ($L_{\text{EX}} = 0$):
\begin{equation}
  d_{\text{harm}} = 10^{(14 - 47.98 + 75)/20} = 10^{2.051} \approx \mathbf{112\text{\,m}}.
  \label{eq:d_harm_num}
\end{equation}

The ratio $d_{\text{LBT}}/d_{\text{harm}} = 685/112 \approx 6.1\times$ confirms that the VLP device defers well beyond its maximum harm range; under SSB-only loading the margin is still $545/112 \approx 4.9\times$. This holds under the \emph{most conservative} model pairing---realistic urban NLoS for detection and optimistic free space for harm; a consistent NLoS treatment of the harm path would shorten $d_{\text{harm}}$ further and widen the margin. LBT alone therefore guarantees spatial protection of both 5G~NR receivers. This margin is referenced to the measured $-75$\,dBm degradation onset; under the more stringent ITU-R/ECC 366 protection criterion of $I/N = -6$\,dB (Table~8 of~\cite{b9}) the corresponding harm distance is larger.

\section{Experimental Setup and Test Approach} \label{sec_experimental_setup}
\label{sec:exp_setup}

The gNB and UE run on the OpenAirInterface5G (OAI) platform~\cite{b12}, extended at ICS, University of Surrey. The following subsections detail the testbed hardware, radio configurations, and measurement procedure.

\subsection{Testbed Overview} \label{sec:testbed}

\begin{figure}[!t]
\begin{center}
\includegraphics[width=1\columnwidth]{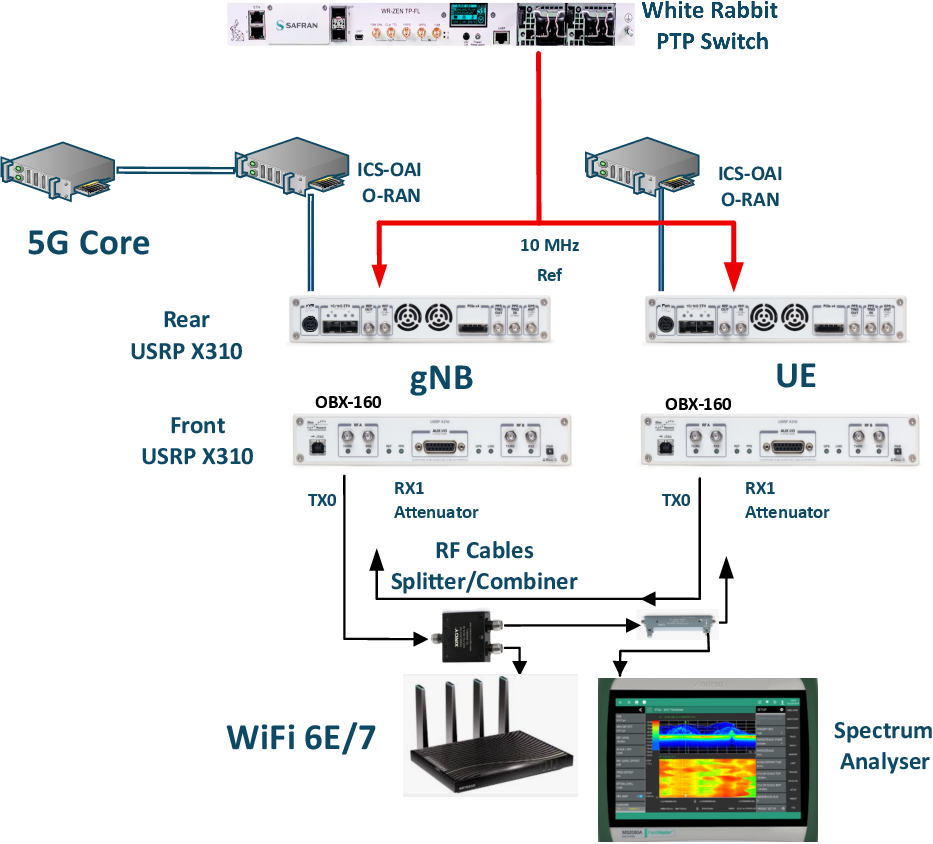}
\end{center}
\caption{5G NR and Wi-Fi 6E/7 Coexistence Testbed}
\label{fig0:testbed}
\end{figure}

Fig.~\ref{fig0:testbed} shows the testbed. Three bare-metal servers host the 5G core, the ICS/OAI~gNB, and the ICS/OAI~UE. Each SDR node uses a USRP~X310 connected via fibre cables to an OBX-160 RF front-end, synchronised via a White Rabbit PTP switch distributing a common 10\,MHz reference. Transmit and receive ports pass through programmable RF attenuators into a ZN4PD1-842-S+ passive combiner that aggregates the 5G~NR DL/UL paths with the Wi-Fi~6E/7 access point (TP-Link AXE5400). A spectrum analyser at the combiner output verifies power levels. The fully cabled architecture gives repeatable control of power and timing without an over-the-air licence.



\subsection{5G NR Configuration} \label{sec:nr_conf}

Tables~\ref{tab:coexistence_params} and~\ref{tab:nr_conf} list the system and gNB-specific parameters. The carrier sits at 5995.2\,MHz in band~n102: 30\,kHz SCS, 106~RBs (40\,MHz), Time Division Duplexing (TDD) pattern of 5\,ms (7~DL slots + 6~DL symbols, 2~UL slots + 4~UL symbols). The ICS/OAI~UE shares the same configuration and transmits at up to 20\,dBm.

\begin{table}[!t]
\centering
\caption{ICS/OAI~gNB and ICS/OAI~UE configurations (band~n102)}
\label{tab:nr_conf}
\footnotesize
\renewcommand{\arraystretch}{1.2}
\begin{tabular}{|l|p{2.2cm}|p{3.2cm}|}
\hline
\textbf{Category} & \textbf{Parameter} & \textbf{Value / Description} \\
\hline
\multirow{5}{*}{Carrier} 
  & DL centre frequency        & 5995.2\,MHz \\ \cline{2-3}
  & UL centre frequency        & 5995.2\,MHz (TDD: same as DL) \\ \cline{2-3}
  & NR band                    & n102 (6\,GHz shared band) \\ \cline{2-3}
  & ARFCN (SSB)                & 799680 \\ \cline{2-3}
  & Point A (ARFCN)            & 798408 (5976.12\,MHz) \\ 
\hline
Waveform
  & Subcarrier spacing         & 30\,kHz ($\mu=1$) \\ 
\hline
\multirow{2}{*}{Bandwidth}
  & Carrier bandwidth          & 106 RBs ($\approx$40\,MHz) \\ \cline{2-3}
  & BWP / bandwidth location   & 28875 (RBstart=0, L=106) \\ 
\hline
\multirow{2}{*}{UL TX}
  & pMax (UE)                  & 20\,dBm (max UE TX power) \\ \cline{2-3}
  & PRB blacklist              & 51, 52, 53, 54 (excluded RBs) \\ 
\hline
TDD
  & Pattern                    & 5\,ms period: 7 DL slots + 6 DL sym, 2 UL slots + 4 UL sym \\ 
\hline
\end{tabular}
\end{table}

\subsection{Wi-Fi 6E Configuration} \label{sec:wifi_conf}

The Wi-Fi~6E access point (TP-Link AXE5400) operates in VLP mode on Channel~9 with an 80\,MHz bandwidth centred at 5985\,MHz. The primary channel frequency is 5995\,MHz, overlapping directly with the 5G~NR carrier. One 6\,GHz U.FL antenna connector is cabled to the coupling port of a directional coupler for conducted injection into the testbed. During each interference measurement, heavy bidirectional traffic is generated between the AP and an Intel~AX210 client adapter by streaming three 8K video sessions and downloading six 10\,Gbit files simultaneously. This load produces sustained channel occupancy throughout the measurement interval.

\subsection{Test Procedure} \label{sec:test_proc}

The Wi-Fi signal is injected into two victim paths:
\begin{itemize}
    \item \textbf{C1 (VLP $\rightarrow$ UE DL):} interference enters the ICS/OAI~UE downlink receiver.
    \item \textbf{C2 (VLP $\rightarrow$ gNB UL):} interference enters the ICS/OAI~gNB uplink receiver.
\end{itemize}

Each path follows a three-step protocol. First, a baseline measurement~(B0) records Reference Signal Received Power (RSRP), Reference Signal Received Quality (RSRQ), Signal to Noise Ratio (SNR), Block Error Rate (BLER), and DL/UL throughput with no interference present. Second, the programmable attenuator is swept in 5\,dB steps to increase the injected Wi-Fi power progressively; finer 1--2\,dB steps are applied near observed thresholds. Third, the same metrics are recorded at each power level. Table~\ref{table:scenario} summarises the test scenarios.

\begin{table}[!t]
\centering
\caption{Measurement scenarios}
\label{table:scenario}
\footnotesize
\renewcommand{\arraystretch}{1.2}
\setlength{\tabcolsep}{2pt}
\begin{tabular}{|c|p{2.2cm}|p{2.0cm}|c|p{1.5cm}|}
\hline
\textbf{ID} & \textbf{Interference} & \textbf{Victim} & \textbf{Atten.} & \textbf{Metrics} \\
\hline
B0 & Off & N/A & -- & RSRP, RSRQ, SINR, TP \\
\hline
C1.1a & Wi-Fi 6E (80\,MHz, traffic) & UE DL (85\,Mbit/s) & 5\,dB & TP, BLER \\
\hline
C1.1b & Wi-Fi 6E (80\,MHz, traffic) & UE DL (60\,Mbit/s) & 5\,dB & TP, BLER \\
\hline
C1.1c & Wi-Fi 6E (beacons only) & UE DL (85\,Mbit/s) & 5\,dB & BLER \\
\hline
C2.1 & Wi-Fi 6E (80\,MHz, traffic) & gNB UL (85\,Mbit/s) & 5\,dB & TP, BLER, SNR \\
\hline
\end{tabular}
\end{table}

TCP DL traffic is generated using \texttt{iperf3} for each test case from the User Plane Function module of the 5G Core towards the UE. Three data sources are captured at every power level:
\begin{itemize}
    \item \textbf{ICS/OAI~gNB logs:} timestamped RSRP, RSRQ, SINR, and BLER.
    \item \textbf{ICS/OAI~UE logs:} \texttt{iperf3} client/server output reporting TCP throughput.
    \item \textbf{Spectrum analyser:} power spectral density captures over the shared 6\,GHz band.
\end{itemize}

\section{Measurement results} \label{sec_measurements}

\begin{figure}[!t]
\begin{center}
\includegraphics[width=0.9\columnwidth]{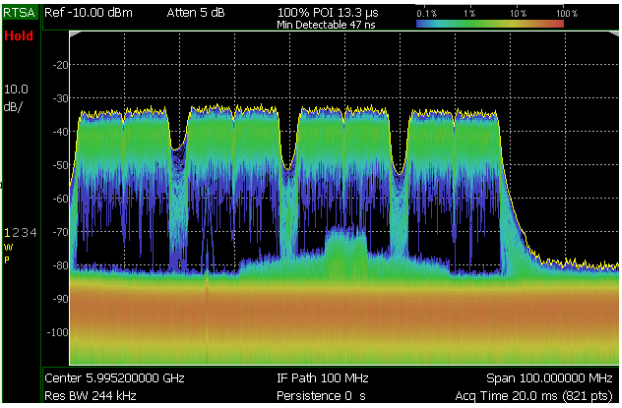}
\end{center}
\caption{Wi-Fi~6E channel~9 (80\,MHz) overlapping NR-ARFCN~799680.}
\label{fig7:channel_overlap}
\end{figure}

This section presents the conducted interference measurements for both victim paths defined in Section~\ref{sec:test_proc}. Fig.~\ref{fig7:channel_overlap} confirms that the Wi-Fi~6E channel overlaps the 5G~NR carrier completely. Tests~C1.1a and C1.1b evaluate two DL operating points. The 85\,Mbit/s rate represents a high-load condition near peak MCS. The 60\,Mbit/s rate represents a medium-load condition with a lower MCS and greater coding margin. Together they cover the range of interference resilience expected in practice.

\subsection{C1.1a: VLP interference to ICS/OAI~UE at 85\,Mbit/s}

The ICS/OAI~UE DL traffic was set to 85\,Mbit/s. With no interference, the measured throughput at the UE matched the configured rate.

\begin{figure}[!t]
\begin{center}
\includegraphics[width=0.9\columnwidth]{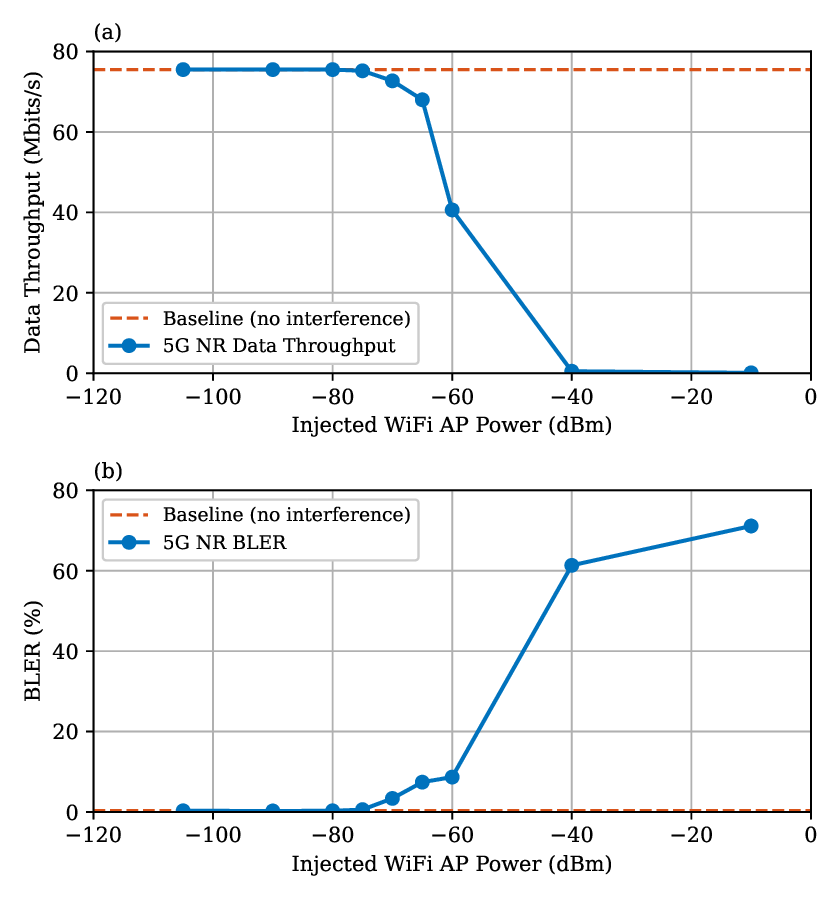}
\end{center}
\caption{ICS/OAI~UE performance versus injected Wi-Fi power (85\,Mbit/s): (a)~DL throughput, (b)~BLER.}
\label{fig_ue85}
\end{figure}

Fig.~\ref{fig_ue85}(a) shows the DL throughput versus injected Wi-Fi power. Throughput was unaffected below $-75$\,dBm. It dropped by ${\sim}10$\% at $-70$\,dBm, ${\sim}25$\% at $-65$\,dBm, and ${\sim}50$\% at $-60$\,dBm. Radio-link failures were observed at $-60$\,dBm. The link became unusable by $-40$\,dBm.

The corresponding BLER is plotted in Fig.~\ref{fig_ue85}(b). BLER remained below 1\% for interferer levels under $-75$\,dBm. It increased to 3.79\% at $-70$\,dBm and reached ${\sim}10$\% at $-60$\,dBm.

\subsection{C1.1b: VLP interference to ICS/OAI~UE at 60\,Mbit/s}

The ICS/OAI~UE DL traffic was set to 60\,Mbit/s. Wi-Fi traffic was unchanged from test~C1.1a. The ICS/OAI~UE was more resilient at the lower data rate. Throughput remained unchanged for interferer levels below $-70$\,dBm and dropped by only 8.58\% at $-60$\,dBm.

BLER, shown in Table~\ref{tab:bler_60}, remained below 1\% for interferer levels up to $-80$\,dBm. It rose to 5.39\% at $-70$\,dBm and reached 9.69\% at $-60$\,dBm.

\begin{table}[!t]
\centering
\caption{ICS/OAI~UE BLER versus injected Wi-Fi power (60\,Mbit/s)}
\label{tab:bler_60}
\footnotesize
\begin{tabular}{|c|c|}
\hline
\textbf{Wi-Fi Power (dBm)} & \textbf{Average BLER (\%)} \\
\hline
$-105$ & 0.29 \\
\hline
$-80$ & 0.06 \\
\hline
$-70$ & 5.39 \\
\hline
$-60$ & 9.69 \\
\hline
\end{tabular}
\end{table}

\subsection{C1.1c: VLP beacon-only interference to ICS/OAI~UE at 85\,Mbit/s}

The ICS/OAI~UE DL traffic was set to 85\,Mbit/s. The Wi-Fi AP transmitted beacons only, with no data traffic. The ICS/OAI~UE was substantially more resilient under beacon-only interference. The average BLER increase was marginal, rising from 0\% to ${\sim}0.05$\% across a 30\,dB increase in interference power (see Table \ref{tab:bler_notraffic}). No significant impact was observed up to $-55$\,dBm.

\begin{table}[!t]
\centering
\caption{Average BLER under beacon-only Wi-Fi interference (no traffic)}
\label{tab:bler_notraffic}
\begin{tabular}{|c|c|}
\hline
\textbf{Wi-Fi Interferer Power (dBm)} & \textbf{Average BLER (\%)} \\
\hline
$-85$ & 0.00 \\
\hline
$-65$ & 0.02 \\
\hline
$-55$ & 0.05 \\
\hline
\end{tabular}
\end{table}

\subsection{C2.1: VLP interference to ICS/OAI~gNB UL receiver}

\begin{figure}[!t]
\begin{center}
\includegraphics[width=0.9\columnwidth]{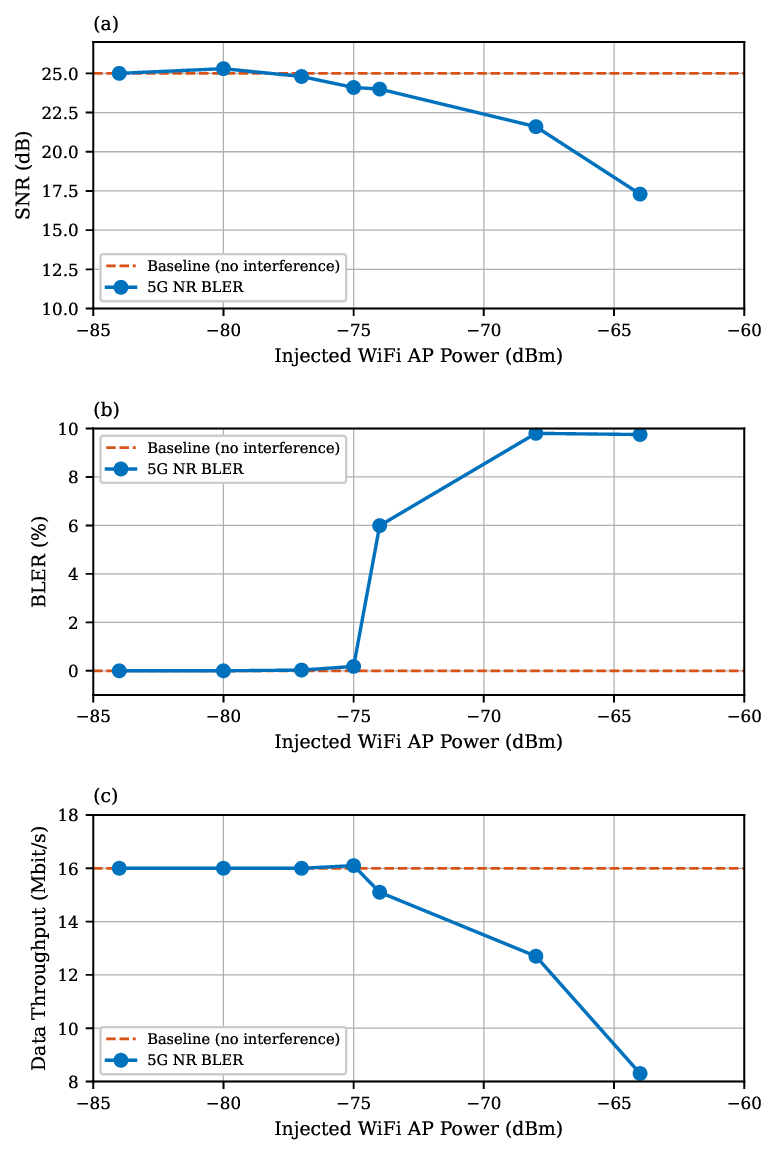}
\end{center}
\caption{ICS/OAI~gNB performance versus injected Wi-Fi power: (a)~mean SNR, (b)~BLER, (c)~UL throughput.}
\label{fig_gnb}
\end{figure}

Wi-Fi interference was injected into the ICS/OAI~gNB UL receiver at progressively increasing power levels. Fig.~\ref{fig_gnb}(a) shows the mean SNR versus injected Wi-Fi power. A 1\,dB SNR drop occurred between $-80$ and $-75$\,dBm. A further 1\,dB drop was measured at $-74$\,dBm, and an additional 6\,dB degradation at $-64$\,dBm.

As shown in Fig.~\ref{fig_gnb}(b), BLER remained below 1\% for interferer levels under $-75$\,dBm. It increased to 10\% at $-64$\,dBm. Fig.~\ref{fig_gnb}(c) shows the UL throughput. No measurable impact was observed below $-75$\,dBm. A 50\% throughput drop occurred at $-64$\,dBm.

\section{Discussion and design implications} \label{sec_discussion}

This section maps the measured degradation thresholds to a real deployment where the VLP device interferes with the 5G~NR gNB and UE receivers. In our conducted tests we bypassed the VLP's LBT and injected Wi-Fi power directly into the victim receiver chain at controlled levels. Table~\ref{tab:impact_summary} summarises the observed impact across all test configurations.

\begin{table}[!t]
\centering
\caption{Summary of VLP Wi-Fi~6E/7 interference impact on 5G~NR receivers}
\label{tab:impact_summary}
\scriptsize
\renewcommand{\arraystretch}{1.15}
\setlength{\tabcolsep}{2pt}
\begin{tabular}{|l|c|c|c|}
\hline
\textbf{Victim} & \textbf{$<-75$\,dBm} & \textbf{$-75$ to $-65$\,dBm} & \textbf{$>-60$\,dBm} \\
\hline
UE (85\,Mbit/s) & No impact & 10--25\% TP loss & 50\%+ TP loss, RLF \\
\hline
UE (60\,Mbit/s) & No impact & 5.4\% BLER & 8.6\% TP loss, 9.7\% BLER \\
\hline
gNB UL & No impact & 1--2\,dB SNR drop & 6\,dB SNR drop, 50\% TP \\
\hline
\multicolumn{4}{|l|}{\textit{Beacon-only interference (no Wi-Fi traffic):}} \\
\hline
UE (85\,Mbit/s) & \multicolumn{3}{c|}{No impact up to $-55$\,dBm (BLER $<0.05$\%)} \\
\hline
\end{tabular}
\vspace{1mm}
\newline
\raggedright
{\scriptsize TP = Throughput; RLF = Radio Link Failure.}
\end{table}

The 5G~NR gNB and UE signals exceed the $-62$\,dBm ED threshold across a radius of 545--685\,m (Section~\ref{subsec:lbt_zone}), so the VLP's own LBT prevents it from transmitting at any distance where its emissions could reach the degradation levels observed in our measurements.

\subsection{VLP interference into the gNB UL receiver}
\label{subsec:gnb_ul}
Applying the path-loss model of Equation~(1) to the $-75$\,dBm gNB UL degradation onset measured in Section~V-D, a VLP device at 14\,dBm EIRP would need to be within 112\,m of the receiver to reach that threshold at its input. At that distance, LBT has already been triggered by the gNB's own transmissions (exclusion zone $\geq 545$\,m). Under normal network operation, the VLP would therefore vacate the channel before its emissions could reach the measured degradation threshold.

\subsection{VLP interference into the UE DL receiver}
Two DL configurations were tested: 85\,Mbit/s (near peak MCS) and 60\,Mbit/s (lower MCS with greater coding margin). The 60\,Mbit/s link was consistently more resilient, and the difference is explained by the lower MCS: more coding redundancy absorbs interference that would push the higher MCS past its BLER target. Under beacon-only interference (test~C1.1c), the ICS/OAI~UE showed no measurable degradation up to $-55$\,dBm, since the low duty cycle of beacon frames poses negligible risk to the 5G~NR DL.

\subsection{Spatial protection through LBT}
The link-budget model of Section~III maps the measured $-75$\,dBm threshold to a VLP harm radius of 112\,m (Equation~(5))---$6.1\times$ shorter than the 685\,m LBT exclusion zone (Equation~(3)). The gNB's LBT contour exceeds the inter-site distance, so both UL and DL symbols remain above the $-62$\,dBm ED threshold throughout the cell. A compliant VLP device cannot transmit on band~n102 anywhere within the network's operational footprint. Study~C10 of ECC Report~366~\cite{b9} reaches the same conclusion through simulation: a single VLP at 14\,dBm does not harm MFCN DL or UL receivers. Our measured onset threshold and 112\,m harm radius provide the first hardware-based confirmation of that finding.
The $-62$\,dBm ED threshold is sufficient for band~n102 deployments. Even at minimum loading (SSB only), the exclusion zone exceeds the harm radius by $4.9\times$. Near an active terminal, a UE transmitting at 23\,dBm EIRP (reduced by 4\,dB body loss~\cite{b9}) triggers the VLP's LBT at approximately 43\,m, providing a secondary protection mechanism. No additional mitigation is needed for a single VLP device. This margin holds as long as the gNB antenna maintains line-of-sight to street level.

\subsection{Limitations}

Bypassing the VLP's LBT and injecting interference continuously represents a worst case that a compliant device would never produce. The testbed uses a single VLP source on a static line-of-sight path; multipath, fading, aggregate interference from co-located devices, adjacent-channel leakage, and mobility are all absent. Real-world interference would therefore be lower, since LBT silencing, fading, and the VLP's 3--5\% duty cycle further reduce exposure.

\section{Conclusion}

This work experimentally characterised co-channel VLP Wi-Fi~6E/7 interference into a 5G~NR system in band~n102 using conducted power-injection measurements. Neither the UE DL receiver nor the gNB UL receiver degraded below $-75$\,dBm, and beacon-only interference caused no impact up to $-55$\,dBm. Mapping these thresholds through the proposed link-budget model, a compliant VLP device vacates the channel before its emissions can degrade either receiver.
Future work will extend these results to adjacent-channel interference, aggregate multi-device scenarios, and band~n104. We also plan to propose mitigation techniques that reduce the impact at the identified degradation threshold.

\end{document}